\documentclass[prodmode]{acmsmall} % Aptara syntax

\usepackage[ruled]{algorithm2e}
\usepackage{multirow}
\usepackage{graphicx}
\usepackage[numbers,comma]{natbib}
\begin{document}

% Page heads
%\markboth{}{Mobile Augmented Reality Survey: A Bottom-up Approach}

% Title portion
\title{Mobile Augmented Reality Survey: A Bottom-up Approach}

\author{ZHANPENG HUANG
\affil{Hong Kong University of Science and Technology, Hong Kong}
PAN HUI
\affil{Hong Kong University of Science and Technology, Hong Kong \& Telekom Innovation Laboratories, Germany}
CHRISTOPH PEYLO
\affil{Telekom Innovation Laboratories, Germany}
DIMITRIS CHATZOPOULOS
\affil{Hong Kong University of Science and Technology, Hong Kong}}
% NOTE! Affiliations placed here should be for the institution where the
%       BULK of the research was done. If the author has gone to a new
%       institution, before publication, the (above) affiliation should NOT be changed.
%       The authors 'current' address may be given in the "Author's addresses:" block (below).
%       So for example, Mr. Abdelzaher, the bulk of the research was done at UIUC, and he is
%       currently affiliated with NASA.

\begin{abstract}
Augmented Reality (AR) is becoming mobile. Mobile devices have many constraints but also rich new features that traditional desktop computers do not have. There are several survey papers on AR, but none is dedicated to Mobile Augmented Reality (MAR). Our work serves the purpose of closing this gap. The contents are organized with a bottom-up approach. We first present the state-of-the-art in system components including hardware platforms, software frameworks and display devices, follows with enabling technologies such as tracking and data management. We then survey the latest technologies and methods to improve run-time performance and energy efficiency for practical implementation. On top of these, we further introduce the application fields and several typical MAR applications. Finally we conclude the survey with several challenge problems, which are under exploration and require great research efforts in the future.
\end{abstract}




%\acmformat{Zhanpeng Huang, Pan Hui, Christoph Peylo and Dimitris Chatzopoulos, 2013. Mobile Augmented Reality Survey: A Bottom-up Approach.}
% At a minimum you need to supply the author names, year and a title.
% IMPORTANT:
% Full first names whenever they are known, surname last, followed by a period.
% In the case of two authors, 'and' is placed between them.
% In the case of three or more authors, the serial comma is used, that is, all author names
% except the last one but including the penultimate author's name are followed by a comma,
% and then 'and' is placed before the final author's name.
% If only first and middle initials are known, then each initial
% is followed by a period and they are separated by a space.
% The remaining information (journal title, volume, article number, date, etc.) is 'auto-generated'.

%\begin{bottomstuff}
%This work is supported by the National Science Foundation, under
%grant CNS-0435060, grant CCR-0325197 and grant EN-CS-0329609.

%Author's addresses: Z. P. Huang, P. Hui, and D. Chatzopoulos, Department of Computer Science and Engineering, Hong Kong University of Science and %Technology, Hong Kong; C. Peylo, Telekom Innovation Laboratories, Germany.
%(Current address) NASA Ames Research Center, Moffett Field, California 94035.
%\end{bottomstuff}

\maketitle
\section{Introduction}
With rapid advances in user interface and mobile computing, comes the ability to create new experience that enhances the way we acquire, interact and display information within the world that surrounds us. Mobile technology improvements in built-in camera, sensors, computational resources and power of cloud-sourced information have made AR possible on mobile devices. We are able to blend information from our senses and mobile devices in myriad ways that were not possible before. Additionally, cloud infrastructure and service providers continue to deploy innovative services to breed new MAR applications. It was estimated \cite{Juniper09} that the MAR-based mobile apps and mobile advertising market is up to \$732 million by 2014.\par
Renevier \cite{Renevier01} and Kock \cite{Kock10} gave their definitions of MAR respectively, but their definitions did not figure out MAR features or limited virtual contents to dynamic 3D objects. However, virtual content can be of any form and presentation as long as it does not exist in the real world through our senses. We extend Azuma's \cite{Azuma01} definition of AR to MAR in a more general way as follows:
\begin{itemize}
    \item Combine real and virtual objects in a real environment.
    \item Interactive in real time.
    \item Register (align) real and virtual objects with each other.
    \item Run and/or display on a mobile device.
\end{itemize}\par
We do not restrict any specified technology to implement a MAR system. Any system with all above characteristics can be regarded as a MAR system. A successful MAR system should enable users to focus on application rather than its implementation \cite{Papagiannakis08}. MAR is the special implementation of AR on mobile devices. Due to specific mobile platform requirements, MAR suffers from additional problems such as computational power and energy limitations. It is usually required to be self-contained so as to work in unknown environments. AR and MAR supplement real world with computer-generated virtual contents other than replace real world, so they are both particulary suitable for scenarios that users require informational support for a task while still need to focus on that task. A typical MAR system comprises mobile computing platforms, software frameworks, detection and tracking support, display , wireless communication, and data management. We will investigate MAR in terms of these aspects in the following sections.\par
\section{System components}
% Head 2
\subsection{Mobile Computing Platforms}
The high mobility of MAR requires it to provide services without constraining users' whereabouts to a limited space, which needs mobile platforms to be small and light. Recent years, we have seen significant progresses in miniaturization and performance improvement of mobile computing platforms.\par
\subsubsection{Notebook computers}
Notebook computers were usually used in early MAR prototypes \cite{Kalkusch02, Cheok03, Piekarski04} as backpack computing platforms. Comparing to consumable desktop computers, notebook computers are more flexible to take alongside. However, size and weight is still the hurdle for wide acceptance by most users. Since notebook computers are configured as backpack setup, additional display devices such as head mounted displays (HMDs) are required for display. Notebook screen is only used for profiling and debugging use.\par
\subsubsection{Personal Digital Assistants (PDAs)} 	
PDAs were an alternative to notebook computers before emergence of other advanced handheld PCs. Several MAR applications \cite{Pasman03, Goose04, Pintaric05, Reitmayr06, Mantyjarvi06, Barakonyi06, Herbst08} configured PDAs as mobile computing platforms, while others \cite{Gausemeier02, Gausemeier03, Wagner03} outsourced the CPU-intensive computations on remote servers to shrink PDAs as thin clients for interaction and display only. PDAs have problems of poor computational capability and absence of floating-point support. Small screen also limits the view angle and display resolution.\par
\subsubsection{Tablet personal computers (Tablet PCs)}
Tablet PC is a personal mobile computer product running Windows operating systems. Large screen size and multi-touch technology facilitate content display and interactive operations. Many MAR systems \cite{Renevier04, Ferdinand05, Guven06, Schmalstieg07a, Quarles08, Han12} were built on Tablet PCs. In addition to expensive cost, Tablet PCs are also too heavyweight for long-time single-handed hold.\cite{Pintaric05}\par
\subsubsection{Ultra mobile PCs (UMPCs)}
	UMPCs have been used for several MAR applications \cite{Newman06, ITACITUS08, Kang08, Tokusho09, Zollner09, Kruijff10}. UMPCs are powerful enough to meet computing requirements for most MAR systems, but they are designed for commercial business market and high price impedes their widespread. Besides, They have similar problems as PDAs due to limited screen size.\par
\subsubsection{Mobile phones}
	Since the first MAR prototype on mobile phone \cite{Mohring04}, mobile phones have achieved great progresses in all aspects from imbedded cameras, built-in sensors to powerful processors and dedicated graphics hardware. Embedded camera is suitable for video see-through MAR display. Built-in sensors also facilitate pose tracking. Many MAR applications \cite{Henrysson04, Henrysson05a, Henrysson05b, Greene06, Rohs07, Schmalstieg08, Takacs08, Morrison09, Chen09, Paucher10, Morrison11, Radha12, Langlotz12} were built on mobile phones. Mobile phones have become predominant platforms for MAR systems because of minimal intrusion, social acceptance and high portability. However, despite rapid advances in mobile phones as computing platforms, their performance for real-time applications is limited. The computing power is still equivalent to a typical desktop computer ten years ago \cite{Paucher10}. Most mobile phones are equipped with relatively slow memory access and tiny caches. Build-in camera has limitations of narrow field-of-view and white noise. Accelerometer sensors are too noisy to obtain accurate position. Magnetometers are also prone to be distorted by environmental factors.\par
\subsubsection{AR glasses}
AR glasses leverage the latest advances in mobile computing and projection display to bring MAR to a new level. They supply a hands-free experience with lest device intrusion. AR glasses work in a nature way that users do not have to look down at mobile devices. However, there is controversy whether they are real MAR or not as current applications on AR glasses supply functions which are irrelative to real world contents and require no tracking and alignment. We regard it as MAR because facial recognition and path finding are suitable for AR glasses and will emerge on AR glasses in near future. In addition, from a much general point of view, AR glasses work as user interface to interact with real world analogous to most MAR systems. Currently the most promising AR glasses is Google Glass \cite{GoogleGlass}. It supports features including taking picture, searching, sending message and giving directions. It is reported to be widely delivered later this year, whereas high price and absence of significant applications may hinder its popularity to some extent.\par
	\vspace{2ex} Figure 1 shows some typical mobile devices used in MAR applications. Since most computing platforms are not designed for MAR use, specific tradeoffs between sizes, weights, computing capability and cost are made for different users and markets. Table~\ref{tab:one} compares mobile platforms in terms of features that MAR may concern.
\begin{figure}
\centerline{\includegraphics[width=13cm]{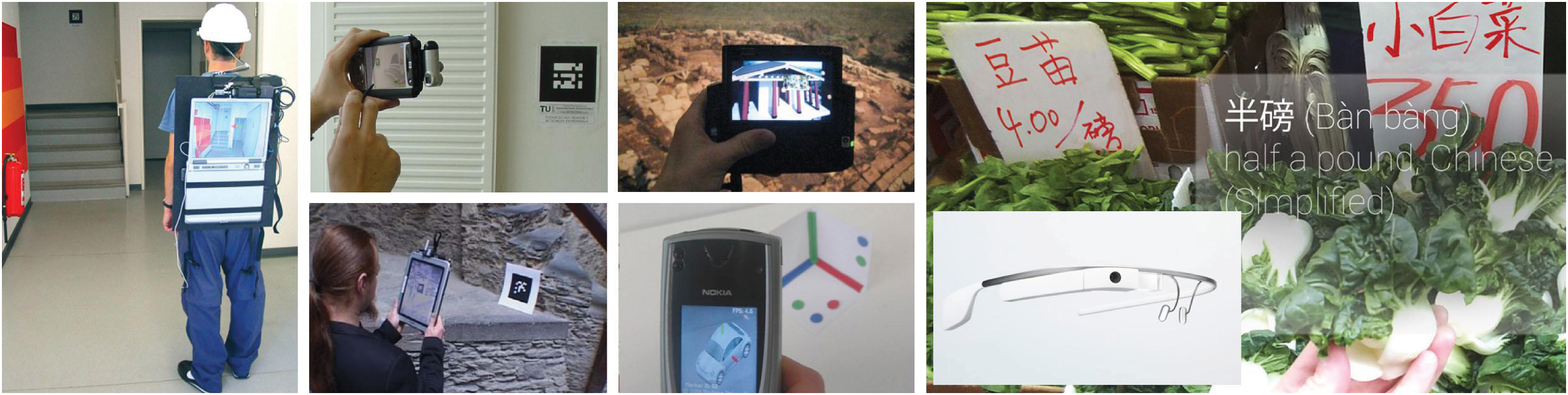}}
\caption{Several mobile devices used in MAR applications. Left: Notebook computer \cite{Reitmayr03a}; Middle: PDA \cite{Wagner03}, Tablet PC \cite{Klein04}, UMPC \cite{Zollner09}, and mobile phone \cite{Mohring04}; Right: Google Glass \cite{GoogleGlass}.}
\label{fig:one}
\end{figure}
\begin{table}%
\tbl{A list of different MAR computing platforms$^a$\label{tab:one}}{%
\begin{tabular*}{\textwidth}{| p{1.25cm} | p{1.4cm} | p{1.5cm} | p{1.8cm} | p{2.3cm} | p{1.3cm} | p{1.4cm} |}
%\begin{tabular*}{\textwidth}{| c | c | c | c | c | c | c | c|}
\hline
Platforms           & Computing Power   & Rendering Power  & Floating-point Support & User Interface  & Portability  & Endurance (hour)      \\\hline
Notebook computer   & high	& high & yes & keyboard/mouse	&low	&2$\sim$4
\\\hline
PDA	                &low	&low   & no  &keyboard/stylus   & high  &4$\sim$6
\\\hline
Tablet PC	        & medium	& medium 	& yes	& stylus/touch	& medium	&5$\sim$8
\\\hline
UMPC	            & medium	& medium	& yes	& keyboard/dialkey/ stylus/touch	& medium	&4$\sim$8
\\\hline
Mobile phone	    & low	  &low	& no$^b$ 	& keyboard /touch	&high	&5$\sim$8
\\\hline
AR glass	        & low	  &medium	&unknown	&voice/touch	&high	&$\sim$6$^c$
\\\hline
\end{tabular*}}
\begin{tabnote}%
\vskip2pt
\tabnoteentry{$^a$}{Characteristics and performance depend on products of different vendors.}
\tabnoteentry{$^b$}{Presently only a few smartphones support floating-point calculation.}
\tabnoteentry{$^c$}{Google Glass.}
\end{tabnote}%
\end{table}%
\subsection{Software Frameworks}
It is complicated and time-consuming to build a MAR system from scratch. Many software frameworks have been developed to help developers focus on high-level applications other than low-level implementations. \par
\subsubsection{Studierstube ES}
	Studierstube \cite{Szalavari98} was developed by the Institute of Computer Graphics in Vienna University of Technology. Reitmayr and Schmalstieg migrated it to mobile platforms as a sub-branch Studierstube ES \cite{Reitmayr01a, Reitmayr01b}. Studierstube ES was rewritten from scratch to leverage newly graphics APIs for better rendering capability. The system supported various display devices and input interfaces. It used OpenTracker \cite{Reitmayr01} to abstract tracking devices and their relations. A network module Muddleware \cite{Wagner07b} was developed to offer fast client-server communication services. A high-level description language Augmented Presentation and Interaction Language (APRL) \cite{Ledermann05} was also provided to author MAR presentation independent of specific applications and hardware platforms. Many MAR applications and prototypes \cite{Pintaric05, Barakonyi06, Schmalstieg07a, Schmalstieg08, Langlotz12} were developed with Studierstube ES. With about two-decade persistent development and maintenance, it became one of most successful MAR frameworks. Currently Studierstube ES is only available for Windows phones and Android platforms.\par
\subsubsection{Nexus}
	Nexus \cite{Hohl99} was developed by University of Stuttgart as a basis for mobile location-aware applications. It supported spatial modeling, network communication and virtual information representation. The architecture was structured in three layers \cite{Nicklas01}, which were used for client devices abstraction, information uniform presentation and basic function wrapper. A hierarchical class schema was dedicated to prepresent different data objects. It supported both local and distributed data management to offer uniform accesses to real and virtual objects. An adaptive module was developed to ensure the scalability in different application scenarios. Nexus prevailed over other platforms in terms of stability and portability.\par
\subsubsection{UMAR}
	UMAR \cite{Henrysson04} was a conceptual software framework based on client-server architecture. It was delicately designed to perform workload as much as possible on the client side to reduce data traffic and over-dependence on network infrastructure. UMAR imported ARToolkit \cite{Kato99} onto mobile phones for visual tracking. A camera calibration module was also integrated to boost accuracy. UMAR was only available to the Symbian platform. Besides, it did not support collaborative MAR applications yet.\par
\subsubsection{Tinmith-evo5}
	Tinmith-evo5 \cite{Piekarski01a, Piekarski03a} was an object-oriented software framework developed by Wearable Computer Lab at the University of South Australia. Data flow was divided into serial layers with sensor data as input and display device as output. All objects in the system were allocated in an object repository to support distributed, persistent storage and run-time configuration. Render system was based on OpenGL and designed to support hardware acceleration. Several MAR applications and games \cite{Piekarski01b, Piekarski02, Piekarski03b} were developed with Tinmith-evo5.\par
\subsubsection{DWARF}
	DWARF \cite{Bauer01, Tonnis03, Bruegge05} was a reconfigurable distributed framework. A task flow engine was designed to manage a sequence of operations that cooperated to finish user tasks. The Common Object Request Broker Architecture (CORBA) was used to construct a peer-to-peer communication infrastructure and manage nodes servers. It was also employed to create wrappers for third-party components. A visual monitoring and debugging tool was developed for fast prototyping. DWARF has been used to create several MAR applications including Pathfinder \cite{Bauer01}, FIXIT for machine maintenance and SHEEP for collaborative game \cite{Bruegge05}.\par
\subsubsection{KHARMA}
	KHARMA \cite{Hill10} was developed by GVU of Georgia Institute of Technology. It was an open architecture based on KML, a type of XML for geo-referenced multimedia description, to leverage ready-to-use protocols and content delivery pipelines for geospatial and relative referencing. The framework contained three major components: channel server to deliver multiple individual channels for virtual content, tracking server to provide location-related information and infrastructure server to deliver information about real world. Irizarry et al. \cite{Irizarry12} developed InfoSPOT system based on KHARMA to access building information. KHARMA supported hybrid multiple sources tracking to increase accuracy, but it was only suitable for geospatial MAR applications.\par
\subsubsection{ALVAR}
	ALVAR \cite{Pasman03} was a client-server based software platform developed by VTT Technical Research Center of Finland. Virtual contents rendering and pose calculation could be outsourced to server to leverage powerful computing and rendering capabilities. Images were then sent back to client and overlaid onto captured images on client for display. It offered high-level tools and methods for AR/MAR developments such as cameral calibration, Kalman filters and markers hiding. ALVAR supported both marker and markerless based tracking as well as multiple markers for pose detection. It was designed to be flexible and independent of any graphical and other third-part libraries except for OpenCV, so it could be easily integrated into any other applications. ALVAR was widely used to construct various MAR applications such as maintenance \cite{Savioja07}, plant lifetime management \cite{Siltanen07} and retail \cite{Valkkynen11}.
\subsubsection{Open source frameworks}
	Besides efforts from academic laboratories, there are several open source frameworks from developer communities. AndAR \cite{andAR} is an open project to enable MAR on Android platforms. The project is still at its early age and it is only tested on very few mobile phones. DroidAR \cite{DroidAR} is similar to AndAR but supports for both location-based MAR and marker-based MAR applications. GRATF \cite{GRATF} is glyph recognition and tracing framework. It provides functions of localization, recognition and pose estimation of optical glyphs in static images and video files. Most open source frameworks are still under development.\par
	\vspace{2ex} Table~\ref{tab:two} gives a comparison of several software frameworks. An ideal software framework should be highly reusable and independent of hardware components and applications. It should be used for various scenarios without reprogramming and modifications. Current frameworks are far from satisfactory for the requirements. It is difficult to abstract all hardware components with a uniform presentation, not to mention that hardware is developing.
\begin{table}%
\tbl{Comparisons of different MAR software frameworks\label{tab:two}}{%
\begin{tabular*}{\textwidth}{|p{1.6cm}|p{1.6cm}|p{1.4cm}|p{2.0cm}|p{2.4cm}|p{2.4cm}|}
\hline
Software    & Programming Language   & Rendering Language    & Auxiliary Tools   & Tracking \& Positioning & Device Support\\\hline
Studierstube ES	&C++	              &OpenGL/ES	          & Authoring tools (APRIL) &ARToolkitPlus \cite{Wagner07a}	&Windows phone / Android mobile devices \\\hline
Nexus	    & unknown	        & unknown	      & AR language (AWML)	    & external sensor system	&Portable computers / handheld devices\\\hline
UMAR	    & Web scripts	    & OpenGL ES$^a$	  & no	                    & ARToolkit \cite{Kato99}	      &Symbian mobile devices\\\hline
Tinmith-evo5 &C++	            & OpenGL	      & no	                    & OpenTracker  \cite{Reitmayr01}	&Portable computer \\\hline
DWARF	    & C++ / Java	    & VRML/ OpenGL	  & Profiling/ debugging tools	&self-contained	        & Portable computers / PDA\\\hline
KHARMA	    & KML\& Web scripts	& OpenGL ES	  & Authoring tools (KML)	& GeoSpots \cite{Hill10}	        & handheld devices \\\hline
ALVAR	    & C++	& third-part graphical libs	  & camera calibration, basic filters	& ARToolkit \cite{Kato99}	        & Portable computers / handheld devices \\\hline
AndAR	    & Java	            & OpenGL	      & no	                    & ARToolkit \cite{Kato99}	    & Android mobile devices  \\\hline
DroidAR	    & Java	            & OpenGL	      & no	                    & self-contained	        & Android mobile devices\\\hline
GRATF	    & C$\sharp$	        & Direct3D	      & Prototyping /debugging tools	        & glyph recognition \cite{Kirillov}	& unknown \\\hline
\end{tabular*}}
\begin{tabnote}%
\vskip2pt
\tabnoteentry{$^a$}{It is uncovered in current version, but it is reported to be OpenGL ES in next version.}
\end{tabnote}%
\end{table}%

\subsection{Display}
\subsubsection{Optical See-through Display}
In optical see-through display, virtual contents are projected onto interface to optically mix with real scene. It requires the interface to be semi-transparent and semi-reflexive so that both real and virtual scenes can be seen. A head tracker is used to obtain users' positions and orientations for contents alignments. Optical see-through display was used in early MAR applications \cite{Feiner99, Renevier01}. It enables users to watch real world with their natural sense of vision without scene distortion. The major problem is that it blocks the amount of light rays from real world and reduces lightness. Besides, it is difficult to distinguish virtual contents from real world when background environment is too bright.
\subsubsection{Video See-through Display}
Video see-through display has two work modalities. One is to use HMD devices to replace user eyes with head-mounted video cameras to capture real world scene. Captured video is blended with computer-generated contents and then sent to HMD screen for display. A head tracker is used to get users position and orientation. This mode is similar to optical see-through display and has been used in early MAR applications \cite{Reitmayr03b, Cheok03, Mohring04}. The other mode works with camera and screen in handheld devices. It uses the embedded cameras to capture live video and blend the video with virtual information before displaying it on the screen. This mode is predominant in applications with handheld devices. The former mode obtains better immersion experience at the cost of less mobility and portability. Comparing to optical see-through display, mixed contents are less affected by surrounding conditions in video see-through display, but it has problems of latency and limited video resolution.
\subsubsection{Surface Projection Display}
Projection display is not suitable for MAR systems due to its consumable volume and high power requirements. With recent progress of projectors in miniaturization and low power consumption, projection display finds its new way in MAR applications. Surface projection displays virtual contents on real object surface rather than display mixed contents on a specific interface. Any object surface, such as wall, paper and even human palm, can be used as interface for display. It is able to generate impressive visual results if real surface and virtual contents are delicately arranged. Pico-projectors have already been used in several MAR applications \cite{Schoning09, Mistry09}. Laser projector, a variation of traditional projector, has been exploited for spatial AR (SAR) applications \cite{Zhou11}. It has many advantages including self-calibrations, high brightness and infinite focal length. Since virtual information is projected to any arbitrary surface, surface projection display requires additional image distortion to match real and virtual projectors for content alignment \cite{project3d}.\par
	\vspace{2ex} Figure 2 illustrates several display devices for MAR. Jannick et al. \cite{Jannick94} gave a detailed comparison of HMD-based optical and video see-through displays in terms of field of view, latency, resolution limitation and social acceptance. Optical see-through display is not often used in recent MAR applications due to sophisticated requirement of projectors and display devices, whereas Google Glass \cite{GoogleGlass} proves that it is also suitable for wearable and MAR systems with micro laser projector.
\begin{figure}
\centerline{\includegraphics[width=9cm]{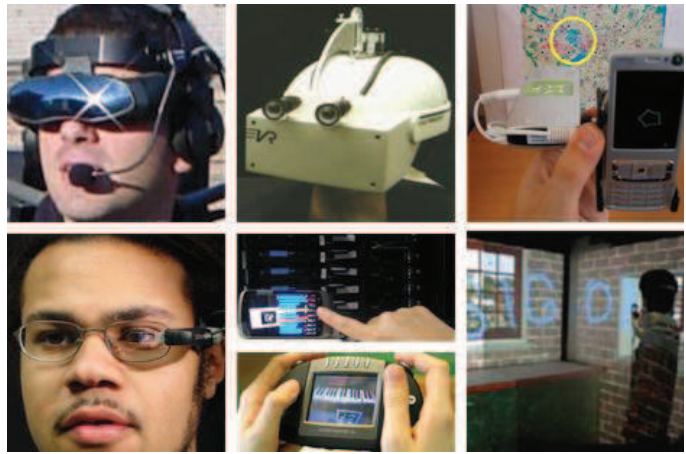}}
\caption{Several display ways in MAR. Optical see-through display (Left): Sony Glasstron LDI-D100B and MicroOptical Clip-on; Video see-through display (Middle): two camera mounted HMD of UNC, mobile phone \cite{Deffeyes11} and PDA \cite{Schmalstieg07a}; Surface project display (Right): mobile camera projector \cite{Schoning09} and laser projector \cite{Zhou11}.}
\label{fig:two}
\end{figure}
\subsection{Tracking and Registration}
Tracking and registration is the process to evaluate current pose information so as to align virtual contents with physical objects in real world. There are two types of tracking and registration: sensor-based and vision-based. Sensor-based methods employ inertial and electromagnetic fields, ultrasonic and radio wave to measure and calculate pose information; vision-based methods estimate gesture information from point correspondent relationships of markers and features from captured images or videos.
\subsubsection{Sensor-based Methods}
According to work modalities, sensor-based methods can be divided into inertial, magnetic, electromagnetic and ultrasonic categories. For simplification, we also categorize inferred-based tracking as a type of electromagnetic method in this paper.\par
	\emph{Inertial-based}: Many inertial sensors output acceleration, which is integrated twice over time to obtain position and angle. Inertial-based method is able to work under most conditions without range limitation or shielding problem. Many MAR applications \cite{Hollerer01, Lang02, Randell03, Hallaway04, White06} used inertial sensors to get user pose information. It has problem of rapid propagation of drift due to double integration and jitters from external interference. Several methods have been proposed to improve accuracy. For example, jitter was suppressed with complementary Kalman filter \cite{Diverdi07}. In \cite{Rolland01}, drift error was minimized by taking relative measurements rather than absolute measurements.The method required a periodic re-calibration and prior knowledge of initial state to get absolute pose in a global reference frame.\par
	\emph{Magnetic-based}: Magnetic tracking uses earth magnetic field to get orientation. It combines with other position tracking methods to obtain six degree of freedom (6DOF). Many MAR applications \cite{Renevier01, Wither05, Hollerer07, Belimpasakis10} used it to track orientation. Magnetic-based method has problem of interference by ambient electromagnetic fields. It is apt to be distorted in surroundings full of metal shields such as steel and concrete skeletons. Chung et al. \cite{Chung11} employed a server that contained magnetic fingerprint map to improve accuracy at indoor area. Sun et al. \cite{Sun12} aggregated ceiling pictures as orientation references to correct original outputs. The method achieved 3.5 times accurate improvement. As it required ceiling patterns for tracking, the method was only suitable for indoor use.\par
	\emph{Electromagnetic-based}: Electromagnetic methods track position based on time of arrivals (TOAs), received signal strength indicator (RSSI) or phase difference of electromagnetic signals. There are several electromagnetic-based tracking methods in MAR field. GPS method is widely used in numerous outdoor MAR applications, but it has problem of single shielding in urban canyons and indoor environments. Besides, result of plain GPS is too coarse to use for accurate tracking. Some works used differential GPS \cite{Dahne02, Reitmayr03b} and real-time kinematic (RTK) GPS \cite{Hollerer01} to improve accuracy but they required locations of base stations. Sen et al. \cite{Sen12} leveraged the difference of Wi-Fi signal attenuation blocked by human body to estimate position. It required communication parameters and  wireless maps beforehand. Wi-Fi tracking has problem of poor accuracy and tedious offline training to construct wireless map \cite{Papagiannakis08}.Ultra Wideband (UWB) is able to obtain centimeter-level accurate results. It has been used in many systems \cite{Kalkusch02, Chung03, Gezici05, Gonzalez07}. The method has drawbacks of high initial implementation cost and low performance. Radio Frequency Identification (RFID) depends on response between RFID readers and RFID tags for tracking \cite{Ni03, Mantyjarvi06, Papagiannakis08}. As each target requires RFID tag for tracking, it is not suitable for massive targets and unknown environments. Infrared tracking works in two ways. One is similar to RFID; the other is to capture infrared beacons as markers with infrared camera. The first way is inexpensive, whereas the second blocks out visual spectrum to provide clean and noise-free images for recognition. Many systems \cite{Goose02, Hallaway03, McFarlane09, Ishiguro10} used both ways for tracking. Infrared signal cannot travel through walls and easily interfere with fluorescent light and direct sunlight. Bluetooth was also used in many applications \cite{Cheok03, Pasman03, Alto04, Rauhala06, Bargh08, Paek10, Wang11}. It is resistable to interference and easier confined to limited space but has drawback of short transmission ranges.\par
\emph{Ultrasonic-based}: Ultrasonic tracking can estimate both pose and velocity information. It is able to obtain very high tracking accuracy. However, ultrasonic emitters and receivers are rarely implemented in nowaday handheld devices. They were only used in early MAR applications \cite{Priyantha00, Foxlin00, Newman01}. Ultrasonic sensors are sensitive to temperature, occlusion and ambient noise and has been replaced by other methods in recent MAR applications.\par
	\vspace{2ex} Table~\ref{tab:three} lists several sensors in terms of characteristics related to MAR. In addition to single sensor tracking, many systems combined different sensors to improve results. In \cite{Cheok03, Hollerer04, Greene06, Schmalstieg07b}, GPS and inertial methods were combined for indoor and outdoors tracking. There are other hybrid methods such as infrared beacons and inertial sensors \cite{Goose02}, UWB and inertial sensors \cite{Kalkusch02}, infrared and RFID \cite{Mantyjarvi06} and Wi-Fi and Bluetooth \cite{Wang11}. We list typical sensor-based tracking methods in Table 2. Several characteristics related to MAR applications are given for comparison.
\begin{table}%
\tbl{A table compares several sensors for tracking in MAR\label{tab:three}}{%
\begin{tabular*}{\textwidth}{|p{1.8cm} |p{2.5cm}| p{2.5cm}| p{2.5cm}| p{2.5cm}|}
\hline
Type                          & Sensors         & Coverage              & Accuracy           & Indoor / outdoor \\\hline
\multirow{2}*{inertial}       & accelerometer	& anywhere	            & 0.01m	                  & indoor / outdoor\\
 \cline{2-5}                  & gyroscopes      & almost anywhere$^a$	& 0.2 deg.	              & indoor / outdoor \\\hline
magnetic	                  & magnetometers	& almost anywhere$^a$	& 0.5 deg.	              & indoor / outdoor\\\hline
\multirow{8}{1.5cm}{electro- magnetic}& GPS	            & almost anywhere$^b$	& 10m$\sim$15m	              & outdoor	\\
 \cline{2-5}                  &differential GPS	& almost anywhere$^b$	& 1m$\sim$3m	                  & outdoor\\
 \cline{2-5}                  & RTK GPS	        & almost anywhere$^b$	    & $\sim$0.01m	              & outdoor	\\
 \cline{2-5}                  & Wi-Fi	        & $\sim$90m	                & 2.5m	                  & indoor\\
 \cline{2-5}                  & RFID	        & 20m$\sim$100m	            & 1m$\sim$3m	                  &indoor\\
 \cline{2-5}                  & UWB	            & 10m$\sim$100m	            & 0.15m$\sim$0.3m	          & indoor\\
 \cline{2-5}                  & infrared	    & $\sim$6m	                & 0.03m$\sim$0.1m	          & indoor\\
 \cline{2-5}                  & Bluetooth	    & $\sim$10m	                & 0.1m$\sim$10m	              & indoor \\\hline
ultrasonic	                  & ultrasonic	    & 10m	                    & $\sim$0.01m	              & indoor \\\hline
\end{tabular*}}
\begin{tabnote}%
\vskip2pt
\tabnoteentry{$^a$}{blind regions including environments full of strong ambient electromagnetic fields and metal shields.}
\tabnoteentry{$^b$}{blind regions including urban canyons, indoor surroundings and places out of LoS.}
\end{tabnote}%
\end{table}%

\subsubsection{Vision-based methods}
Vision-based tracking uses feature correspondences to estimate pose information. According to features it tracks, it can be classified into marker-based and feature-based method.\par
\emph{Marker-based method}: Marker-based method uses fiducials and markers as artificial features. Figure 3 gives several common used markers. A fiducial has predefined geometric and properties, such as shape, size and color patterns, to make them easily identifiable. Planar fiducial is popular attributed to its superior accuracy and robustness in changing lighting conditions. Several available libraries used planar fiducial for tracking, such as ARToolkit \cite{Kato99}, ARTag \cite{Fiala05}, ARToolKitPlus \cite{Wagner07a} and OpenTracker \cite{Reitmayr01}. Many MAR applications \cite{Reitmayr01a, Piekarski03a, Henrysson04, Schmalstieg07a, Schmalstieg08} used these libraries for tracking. Mohring et al. \cite{Mohring04} designed a color-coded 3D paper marker to reduce computation. It could run at 12 fps on a commercial mobile phone. The method was constrained to simple tracking related to 2D position and 1D rotation. Hagbi et al. \cite{Hagbi09} used shape contour concavity of patterns on planar fiducial to obtain projective-invariant signatures so that shape recognition was available from different viewpoints. As it is impractical to deploy and maintain fiducial markers in an unknown or large outdoor environment, marker-based method is only suitable for indoor applications. Besides, it suffers from problems of obtrusive, monotonous and view occlusion.\par
\begin{figure}
\centerline{\includegraphics[width=11cm]{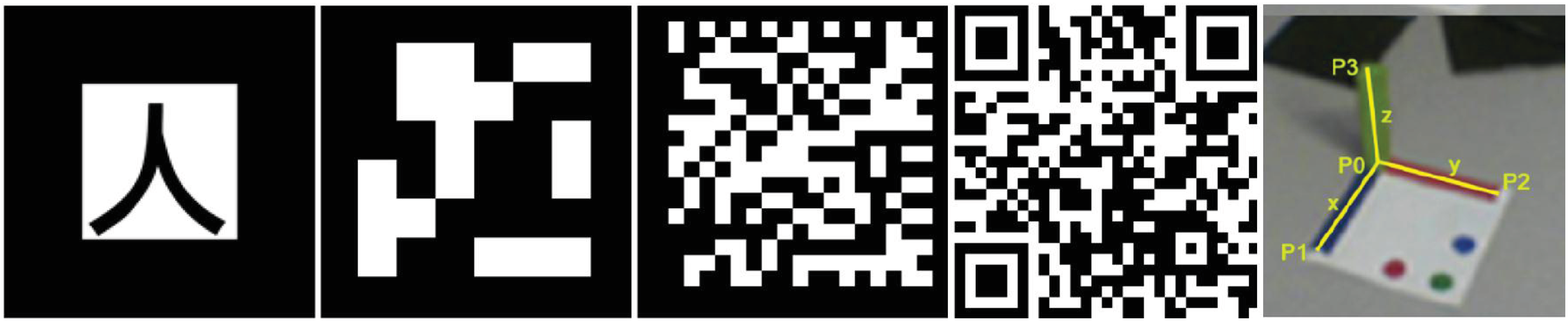}}
\caption{Some visual markers used for tracking. From left to right: template markers \cite{Kato99}, BCH markers \cite{Schmalstieg07a}, DataMatrix markers \cite{Schmalstieg08}, QR barcode markers \cite{Deffeyes11}, and 3D paper marker \cite{Mohring04}.}
\label{fig:three}
\end{figure}\par
	\emph{Nature feature method}: Nature feature method tracks point and region features in image sequences to calculate correspondent relationships to estimate pose information. The method requires no prior information of environment. The frame-by-frame tracking helps to remove mismatches and drift errors that most sensor-based methods suffer. However, it suffers from deficiencies of image distortion, illumination variation and self-occlusion \cite{Neumann99}. Besides, additional registration of image sequence with real world is required to get finally results. \par
	Many robust local descriptors including SIFT \cite{Lowe04} and SURF \cite{Bay06} are introduced for nature feature tracking in MAR field. Skrypnyk and Lowe \cite{Skrypnyk04} presented a traditional SIFT-based implementation. SIFT features were extracted offline from reference images and then used to compute camera pose from live video. Fritz et al. \cite{Fritz06} deployed an imagery database on a remote server and conducted SIFT feature matching on the server to reduce computing and memory overhead on client mobile phones. Chen et al. \cite{Chen09} also implemented the SURF tracking on remote server. The live video captured by embedded camera was streamed to server for tracking. The result was then retrieved back to client side for display. \par
	The major problem of nature feature method is expensive in terms of computational overhead, which is especially severe for MAR applications with requirement of real time performance on mobile devices. Many researches focused on performance improvement from different aspects, such as GPU acceleration, computation outsourcing and algorithm improvement. We focus on algorithm improvement in this section with other two aspects in the next section. Wagner et al. \cite{Wagner08} replaced conventional Difference of Gaussians (DoGs) with FAST corner detector, 4*4 sub-regions with 3*3 sub-regions and k-d tree with multiple Spill Trees to reduce SIFT computation. The reduced SIFT was self-contained and tracked 6DOF at 20Hz frame rate on mobile phones. Takacs et al. \cite{Takacs07} identified new contents in image sequences and conducted feature extraction and database access only when new contents appeared. They further proposed a optimization scheme \cite{Takacs08} to decrease feature matching by rejecting irrelevant candidates according to location information. Only data from nearby location cells were considered. They further developed a feature clustering and pruning strategy to eliminate redundant information. It reduced both computation and bandwidth requirements. An entropy-based compression scheme was used to encode SURF descriptors 7 times faster. Their method run 35\% faster with only half memory consumption. Chen et al. \cite{Chen07} used integral image for Haar transformation to improve SURF computational efficiency. A Gaussian filter lookup table and an efficient arctan approximation were used to reduce floating-point computation. The method achieved performance improvement of 30\% speedup. Ta et al. \cite{Ta09} proposed a space-space image pyramid structure to reduce search space of interest points. It obtained real-time object recognition and tracking with SURF algorithm. In order to speed up database querying, images were organized based on spatial relationships and only related subsets were selected for matching. The method obtained 5 times speedup compared to original SURF and it took less than 0.1s on Nokia N95 mobile phones. Li et al. \cite{Li12} quantized a small number of random projections of SIFT features and sent them to the server, which returned meta-data corresponding to the queried image by using a nearest neighbor search in the random projection space. The method achieved retrieval accuracy up to 95\% with only 2.5kB data transmission. Park et al. \cite{Park12} optimized original feature tracking process from various aspects. Since stable features did not lose tracking until they were out of view, there was no need to extract them from each frame. Feature point number was also limited to 20 to reduce computation. A feature prediction excluded feature points that would disappear and added new features that may appear after a period of time. It achieved runtime performance of 35ms per frame, about 7 times faster than original method. Wagner et al. \cite{Wagner09a} proposed a multiple target tracking and detection method on mobile phones. It separated target detection and tracking so that target tracking run at full frame rate and created a mask to guide detector to look for new target. To further reduce calculations, they assumed that local area around positive matched key points were visible in camera images and uncovered regions were ignored during key point detection. The method was able to track 6 planar targets simultaneously at a rate of 23 fps on an Asus P565 Windows mobile phone.
\subsubsection{Hybrid Tracking Methods}
Each individual method has its advantages and limitations. A better solution is to overcome inherent limitations of individual method by combining different methods together. For instance, inertial sensors are fast and robust under drastic rapid motion. We can couple it with vision-based tracking to provide accurate priors under fast movements. Behringer et al. \cite{Behringer02} proposed a hybrid tracking algorithm to estimate the optimal 3D motion vector from displacements of 2D image features. The hybrid method employed GPS and digital compass to obtain an approximate initial position and orientation. A vision tracking then calculated camera pose by predicting new features in perspective projection of environmental models. It obtained a visual tracking precision of $0.5^o$ and was able to worked under a maximal rotation motion of $40^o/s$. Jiang et al. \cite{Jiang04} used gyroscope to predict orientation and image line positions. The drift was compensated by a line-based vision tracking method. A heuristic control system was integrated to guarantee system robustness by reducing re-projection error to less than 5 pixels after a long-time operation. Hu and Uchimura \cite{Hu06} developed a parameterized model-matching (PMM) algorithm to fuse data from GPS, 3D inertial gyroscope and vision tracking. Inertial sensor was used to evaluate initial motion and stabilize pose output. The method obtained convincing precise and robust results. Reitmayr and Drummond \cite{Reitmayr06} combined vision tracking with gyroscope to get accurate results in real time. Vision tracker was used for localization and gyroscope for fast motion. A Kalman filter was used to fuse both measurements. Honkamaa et al. \cite{Honkamaa07} used GPS position information to download Google Earth KML models, which were aligned to real world using camera pose estimation from feature tracking. The method strongly depended on the access to models on Google server. Paucher and Turk \cite{Paucher10} combined GPS, accelerator and magnetometer to estimate camera pose information. Images out of current view were discarded to reduce database search. A SURF algorithm was then used to match candidate images and live video to refine pose information. Langlotz et al. \cite{Langlotz12} used inertial and magnetic sensors to obtain absolute orientation, and GPS for current user position. A panorama-based visual tracking was then fused with sensor data by using a Kalman filter to improve accuracy and robustness.\par
	\vspace{2ex} Sensor-based method works in an open loop way. Tracking error can not be evaluated and used for further correction. Besides, it suffers from deficiencies of shielding, noise and interference. Vision-based method employs tracking result as feedback to correct error dynamically. It is analogous to closed loop system. However, tracking accuracy is sensitive to view occlusion, clutter and large variation in environment conditions. Besides, single feature-based method is prone to result inconsistence\cite{Irizarry12}. Hybrid tracking requires fusion \cite{Hu06, Reitmayr06, Diverdi07, Langlotz12} of results from different sources. Which method to use depends on accuracy and granularity required for a specific application scenario. For instance, we can accept some tracking error if we annotate the rough outlines of a building, whereas more accurate result is required to pinpoint a particular window in the building.

\subsection{Wireless Networks}
We investigate wireless networks for pose tracking in previous section, but they are also widely used for communication and data transmission in MAR. As fast progresses in wireless network technologies and infrastructure investments, numerous systems were built on client-server architecture by leveraging wireless networks for communication. There are three major wireless networks used in MAR applications: \par
\subsubsection{Wireless Wide Area Network (WWAN)}
WWAN is suitable for applications with large-scale mobility. There are massive WWAN implementations based on different technologies including 2G GSM and CDMA, 2.5G GPRS, 3G UMTS and 4G LTE. Higher generation network usually has much wider bandwidth and shorter latency than lower generation network. Many MAR applications \cite{Henrysson04, Schmalstieg07b, Takacs08, Tokusho09, Wu11}used WWAN for data transition and communications. One problem of WWAN is high cost of initial infrastructures investment. Besides, networks supplied by different providers are incompatible and users usually have to manually switch different networks. However, WWAN is still the most popular, sometimes the only, solution for MAR communication as it is the only available technology for wide public environments at present.\par
\subsubsection{Wireless Local Area Network (WLAN)}
WLAN works in a much smaller scope but with higher bandwidth and lower latency. Wi-Fi and MIMO are two typical WLANs. WLAN has become popular and is suitable for indoor applications. Many MAR applications \cite{Reitmayr03b,Hollerer04,Pintaric05,Chen09} were built on WLAN based architecture. However, limited coverage may constrain it be used in wide public environments.\par
\subsubsection{Wireless Personal Area Network (WPAN)}
WPAN is designed to interconnect devices, such as mobile phones, PDAs and computers, centered around individual workspace. There are many WPAN implementations including Bluetooth, ZigBee and UWB. Bluetooth and ZigBee are usually used for position tracking and data transmission \cite{Henrysson05a, Henrysson05b, Schmalstieg07a, Schoning09}, whereas UWB is major for tracking. WPAN has many advantages including small volume, low power consumption and high bandwidth, but it is not suitable for application with wide whereabouts.\par
	\vspace{2ex} Table~\ref{tab:four} gives comparison of several wireless networks used in MAR applications. All technologies have their drawbacks. We can leverage advantages of different networks to improve performance if multiple wireless networks overlap. However, it requires manually switching between different networks. \cite{McCaffery04} proposed a wireless overlay network concept to choice the most appropriate available networks for use. It was totally transparent for applications and users to switch between different networks.
\begin{table}%
\tbl{A table comparing different types of wireless networks for MAR applications\label{tab:four}}{%
\begin{tabular*}{\textwidth}{|p{1.0cm} |p{1.5cm}| p{2.0cm}| p{3.0cm}| p{2.5cm}| p{1.4cm}|}
\hline
Type                          & Technology         & Coverage       & Bandwidth(bps)      & Latency(ms)      & Power(mw)  \\\hline
\multirow{4}*{WWAN}           & GSM	               & $\sim$35km	            & 60K	                  & high	         & 1000$\sim$2000 \\
 \cline{2-6}                  & CDMA               & 250km$\sim$350km	& 384K	                  & high	         & 200$\sim$100 \\
 \cline{2-6}                  & GPRS               & $\sim$10km         & 56K$\sim$114K           & high             & $\sim$1000 \\
 \cline{2-6}                  & UMTS               & 1km$\sim$2km       & 2M                      & medium           & $\sim$250 \\\hline
\multirow{2}*{WLAN}	          & Wi-Fi	           & $\sim$90m	        & 11M$\sim$54M	          & low	             & $\sim$100 \\
 \cline{2-6} 	              & MIMO	           & $\sim$100m	        & 300M	                  & medium	         & unknown \\\hline
\multirow{3}*{WPAN}           & UWB	               & 10m$\sim$100m	    & 20M$\sim$1G	          & low              & 20$\sim$1000 \\
 \cline{2-6}                  & Bluetooth	       & $\sim$10m	        & 1M$\sim$3M	          & medium	         & 1$\sim$2.5 \\
 \cline{2-6}                  & ZigBee	           & $\sim$75m	        & 20K$\sim$250K	          & low	         & 1$\sim$100 \\\hline
\end{tabular*}}
\begin{tabnote}%
\vskip2pt
\end{tabnote}%
\end{table}%

\subsection{Data Management}
Any practical MAR application requires efficient data management to acquire, organize and store large quantities of data. It is nature to design dedicated data management for specified applications, but it can not be reused and scaled for other applications. We require more flexible strategies to present and manage data source so as to make it available for different applications.\par
\subsubsection{Data Acquisition}
MAR requires a dataset model of user environment which includs geometrical models and semantic description. Many applications create such model manually, whereas scaling it to a wide region is impractical. \par
Data conversion from legacy geographic databases, such as GIS, is a convenient approach. Reitmayr and schmalstieg \cite{Reitmayr03b} extracted geographic information from a network of routes for pedestrians. Schmalstieg et al. \cite{Schmalstieg07b} extracted footprints information of buildings from 2D GIS database. Some works constructed environmental model from architectural plans. Hollerer \cite{Hollerer04} created building structures from 2D map outlines of Columbia campus. As legacy database normally does not contain all necessary information, the method requires knowledge from other fields to complete modeling. \par
Field Survey with telemetry tools is also widely used to obtain environmental geometry data. Joseph et al. \cite{Joseph05} developed a semi-automatic survey system by using fiducial markers to guide a robot for measurement, based on which Schmalstieg \cite{Schmalstieg07b} employed Leica Total Station TPS700 theodolite to scan indoor structure of buildings. The cloud point results were loaded into a view editor for manual construction of floors, walls and other elements. They further used a robot for surveying automatically. Output cloud points could be converted into DFX format using software packages, which was friendly to 3D applications. Results from measurement are prone to inaccuracy and noise. Besides, discrete cloud points require interpolation to rebuild cartographic presentation. \par
Many MAR browsers \cite{Layer, Wikitude} concerned location-based services augmented with web information. They used geo-location information, images and QR markers to search correlated contents through Internet. Results were retrieved and overlaid on current view on mobile phones. Recent products including Google Earth and Semapedia offer geo-referenced geometry data through community efforts, which are easy to access through Internet. Hossmann et al. \cite{Hossmann12} developed application to gather environmental data through users report of their current activities. The environmental data coverage explodes as users increase.
\subsubsection{Data Modeling}
MAR Applications potentially do not access the same abstraction or presentation of dataset even with the same resource. Besides, it is difficult to guarantee presentation consistency if any change cannot be traced back to the original resource. High-level data model is required to decouple underlying resource from upper logic changes. A data model is a conceptual model to hide data details so as to facilitate understanding, representation and manipulation in a uniform manner. \par
	Researchers at Vienna University of Technology proposed a 3-tier \cite{Reitmayr03b, Schmalstieg07b} data model. The first tier was a database. The second tier linked database and application by translating raw data from database to specified data structure. The third tier was a container where all applications resided. The second tier decoupled data from presentation so that applications did not have to understand data details. Basic abstract types such as \textit{ObjectType} and \textit{SpatialObjectType} were predefined that application types could derive from. The XML object tree was interpreted in a geometrical way so that data storage and presentation were linked and consistent. As data were modeled with nodes, it may increase search computation for rendering engine when several information was not required.\par
	Nicklas and Mitschang \cite{Nicklas01} also proposed a 3-layer model including client device layer, federation layer and server layer. The server layer stored resource for entire system. It could be geographical data, users' locations or virtual objects. A top-level object \textit{Nexus Object} was designed, from which all objects such as sensors, spatial objects and event objects could inherite. The federation layer provided transparent data access to upper layer using a register mechanism. It decomposed query from client layer and then dispatched them to registers for information access. The federation layer guaranteed consistent presentation even if data servers supplied inconsistent data. The model separated underlying data operations from client layer, but it increased access delay due to delegation mechanism. Multiple copies of object on different servers caused data inconsistence.\par
	Other than traditional 3-layer structure, Tonnis \cite{Tonnis03} proposed a 4-layer data model. The bottom layer is a dynamic peer-to-peer system to provide basic connectivity and communication services. The second layer supplied general MAR functions such as tracking, sensor management and environmental presentation. The third layer contained high-level functional modules composed of sub-layer components to offer application related functions for top layer that directly interacted with users and third-part systems. Virtual object were represented by object identifier \textit{virtualObjectID} and their types, which were bound to a table data structure \textit{object\_properties} containing linking information. A data structure \textit{object\_relations} was proposed to describe object relationships. A special template was also used to store representative information. The flat and separate data model was more flexible and efficient for different application and rendering requirements.\par
\subsubsection{Data Storage}
	Since no global data repository exists, researchers have to build private data infrastructure for their MAR applications. Hollerer et al. \cite{Hollerer01, Hollerer04} constructed a relational central database to access meta-data based on client-server architecture. A data replication infrastructure was applied to distribute data to various clients. Piekarski et al. \cite{Piekarski01a, Piekarski03a} implemented hierarchical database storage which stored and retrieved objects using hierarchical path names similar to a file system. It used virtual directories to create and store objects without understanding creation details. Reitmayr and Schmalstieg \cite{Reitmayr03a, Reitmayr03b} used a file system to organize application data in their early implementation. As file system is efficient for unstructured dataset whereas MAR data is usually well structured, Schmalstieg et al. \cite{Schmalstieg07a, Schmalstieg07b, Schmalstieg08} adopted a XML database with a XML-based query strategy, which was proven more flexible and efficient. Wagner \cite{Wagner07b} developed a middleware named Muddleware to facilitate database access. It had a server-side state machine to respond to any database changes. An independent thread was dispatched to control database server. Nicklas and Mitschang \cite{Nicklas01} advocated a multi-server infrastructure to decouple different data processes but it had problem of data inconsistence as data copies were deployed on multiple servers. Conventional database technologies were also widely used to store and access various resources in many MAR applications. In terms of massive data storage technologies, from user's point of view, the important issue is how to get to the most relevant information with the least effort and how to minimize information overload.
\section{System Performance and Sustainability}
Most MAR applications suffers from poor computing capability and limited energy supply. To develop applications for practical use we should consider issues of runtime performance and energy efficiency. From the developers' point of view, we should make design and development decisions based on careful task analysis.
\subsection{Runtime Performance}
Recent years we have witnessed great efforts to improve runtime performance for mobile applications. A speedup from hardware, software and rendering improvements, ranging from a few times to hundreds of times, has been achieved during the past decade.
\subsubsection{Multicore CPU parallelism}
There are many off-the-shelf multicore CPU processors available for mobile devices, such as dual-core Apple A6 CPU and quad-core ARM Cortex-A9 CPU. [Softpedia09] reported that about 88\% mobile phones would be equipped with multicore CPU by 2013. Multicore CPU consumes less energy than single core CPU with similar throughput because each core works at much lower clock frequency and voltage. Most MAR applications are composed of several basic tasks including camera access, pose tracking, network communication and rendering. Wagner and Schmalstieg [Wagner09b] parallelized basic tasks for speedup. Since camera access was I/O bound rather than CPU bound, they run camera reading in a separate thread. Herling and Broll [Herling10] leveraged multicore CPUs to accelerate SURF on mobile phones by treating each detected feature independently and assigning different features to different threads. Takacs et al. [Takacs11] separated detection and tracking into different threads for parallelization. The system run at 7$\sim$10 fps on a 600MHz single-core CPU. Multithread technology is not popular for MAR applications as computing context is much more stringent than desktop computers. CPU and other accelerators are integrated into single processor system-on-chip (SoC) to share system bus. It requires intelligent scheme to schedule threads to share data and avoid access conflict.
\subsubsection{GPU Computing}
Most mobile devices are now equipped with mobile graphics processing unit (GPU). There are many mobile GPUs including Qualcomm SnapDragon SoC with Adreno 225 GPU, TI OMAP5 SoC with PowerVR SGX 544 and Nvidia Tegra 3 SoC with ULP GeForce. Mobile GPU is developing toward programmable rendering pipeline. In order to facilitate mobile GPU programming, Khronos Group proposed a low-level graphics programming interface named OpenGL ES \cite{OpenGLES}. It supports per vertex and per pixel operations by using vertex and pixel (or fragment) shaders, which are C-like program code snippets that run on GPUs. The programmability and inherent high parallel architecture awake research interests in general computing acceleration beyond graphics rendering, namely general-purpose computing on GPU (GPGPU). However, it is complicated to program shaders as we have to map algorithms and data structure to graphics operations and data types. To alleviate the problem, Khronos Group released high-level APIs named OpenCL \cite{OpenCL}. They also released the embedded profile of OpenCL for mobile devices.\par
Profile results \cite{Pasman03, Mohring04, Billinghurst08} have shown that tracking is a major performance bottleneck for most MAR applications. Many works accelerated feature tracking and recognition with GPU on mobile devices. Wutte \cite{Wutte09} implemented SURF on hybrid CPU and GPU with OpenGL ES 2.0. Since compression and search processes run on CPU, it required frequent data transmissions between CPU and GPU. Kayombya \cite{Kayombya10} leveraged mobile GPU to accelerate SIFT feature extraction with OpenGL ES 2.0. They broke down the process into pixel-rate sub-process and keypoint-rate sub-process and then projected them as kernels for stream processing on GPU. It took about 0.9s to 100\% match keypoint positions with an image of size 200*200. Singhal et al. \cite{Singhal10, Singhal11} developed several image processing tools including Harris corner detector and SURF for handheld GPUs with OpenGL ES 2.0. They pre-computed neighboring texture coordinates in the vertex shader to avoid dependent texture reading for filtering in fragment shader. Other optimizations such as lower precision and texture compression were used to gain further improvements. Their GPU-based SURF implementation cost about 0.94s for image of size 800*480, about 1.81x speedup comparing to CPU implementation. Hofmann \cite{Hofmann12} also implemented SURF on mobile GPU with OpenGL ES 2.0. The method used mipmaps to obtain scale-aware, sub-pixel-accurate Haar wavelet sampling. It took about 0.4s to extract 1020 SURF-36 descriptors from image of size 512*384 on mobile phones. Leskela et al. \cite{Leskela09} conducted several image processing tests on mobile GPU with OpenCL. The results were inspiring. However, in order to save energy consumption, mobile GPU is designed with low memory bandwidth and a few stream processors (SPs) and instruction set is also reduced.
\subsubsection{Cache Efficiency}
Many mobile devices have tiny on-chip caches around CPU and GPU to reduce latency of external memory access. For instance, Nvidia Tegra series mobile GPUs have a vertex cache for vertex fetching and a pixel and a texture caches for pixel shader. The caches are connected to a L2 cache via system bus. Memory is designed to be small to reduce energy consumption on mobile devices. Cache miss is therefore more expensive than desktop computers. Wagner and Schmalstieg \cite{Wagner09b} transmitted vertex data in interleaving way to improve cache efficiency. They further employed vertex buffer objects (VBOs) for static meshes. Many systems leverage multithreading technologies to hide memory latency. PowerVR SGX5xx GPU \cite{PowerVR11} scheduled operations from a pool of fragments when a fragment waited for texture requests. The method was effective but not energy-efficient. Besides, mobile CPU and GPU only had a few thread wraps due to power constraint, which might not hide cache misses efficiently. Arnau et al. \cite{Arnau12} decoupled pixel and texture access from fragment operations. The architecture fetched data earlier than it would be used so as to hide cache latency when it was used. It achieved 93\% performance of a highly threaded implementation on desktop computer. Hofmann \cite{Hofmann12} employed mipmap technology for Haar sampling. The method obtained a beneficial side effect of significant cache efficiency as it reduced texture fetch operations from 64 to 1. Cache optimization is usually an ad hoc solution for specified problem. For instance, in \cite{Cheng11}, as problem dimension increased, data had to be stored in off-chip memory as on-chip GPU cache was not large enough. Their method was only capable for small dimensional problem. Performance improvement depends on both available hardware and problem to solve. A side benefit of cache saving is to reduce bus accesses and alleviate bandwidth traffic.
\subsubsection{Memory Bandwidth Saving}
Most mobile devices adopt share storage architecture to reduce energy consumption and cost. CPU, GPU and other processors use common system bus to share memory, which makes bandwidth scarce and busy. Besides, bandwidth is designed to be small for low energy cost. \cite{Owens05} reported that annual processor computational capability grew by 71\% whereas bandwidth only by 25\%, so bandwidth is more prone to be bottleneck than computation power. Date compression is usually used to alleviate the problem. Color, depth and stencil buffer data were compressed to reduce transmission on system bus \cite{Moller08, Capin08}. Texture is also compressed before it is stored in constant memory. As slight degradation of image quality is acceptable due to filter and mipmap operations during texture mapping, lossy compressions can be used to further reduce data. Moller and Strom \cite{Moller03, Strom05} proposed a hardware rasterization architecture to compress texture on mobile devices. It reduced memory bandwidth by 53\%. Singhal et al. \cite{Singhal11} stored texture with low precise pixel format such as RGB565, RGBA5551 and RGBA444 to reduce data occupation. Wagner and Schmalstieg \cite{Wagner07a} used native pixel format YUV12 for images captured by built-in cameras to reduce image storage.
\subsubsection{Rendering Improvement}
3D graphics rendering is one of the most computational intensive tasks on mobile device. Duguet and Drettakis \cite{Duguet04} proposed a point-based method to reduce model presentation for cheap 3D rendering on mobile device. They represented model mesh as hierarchical points and rendered parts of them according to computational power of mobile devices. Setlur et al. \cite{Setlur05} observed limitation of human perception for unimportant objects on small display screen. They augmented important objects and eliminated unimportant stuff to reduce rendering. It is an approximate rendering as some information is discarded in final result. Another approximate rendering method is programmable culling unit (PCU) proposed by Hasselgren and Moller \cite{Hasselgren07}. It excluded pixel shader whose contribution to entire block of pixels is smaller than zero. The method obtained 2 times performance improvement with about 14\% to 28\% bandwidth saving. It could degrade to lossy rendering if contribution factor was set to be greater than zero. Traditional immediate mode rendering (IMR) mode updates entire buffer through pipeline immediately. Mobile on-chip memory and cache are too small and rendering data has to be stored in off-chip memory. Fuchs et al. \cite{Fuchs89} proposed a tile-based rendering (TBR) to divide data into subsets. It performed rasterization per tile other than entire frame buffer. As tiny data was required for each tile rendering, data for a tile can be stored in on-chip memory to improve cache and bandwidth efficiency. Imagination Technologies Company delivered several TBR-based PowerVR series GPUs \cite{PowerVR11} for mobile devices. As triangles were required to sort for each tile, additional computation would increase computational cost and bandwidth usage. Performance gain depends on scene geometry structure. It obtains greater performance improvement if overdraw is high, while it may be less efficient than IMR if scene is full of long and thin triangles.
\subsubsection{Computation Outsourcing}
As rapid advances in high bandwidth, low latency and wide deployment of wireless network, it is feasible to outsource computational intensive and even entire workloads to remote server and cloud.\par
\emph{Local-rendering remote-computing (LRRC)}: Computational tasks are outsourced to remote server for acceleration and results are retrieved back to client for rendering and further processes. Chang and Ger \cite{Chang02} proposed an image-based rendering (IBR) method to display complex models on mobile devices. They used a computational intensive ray-tracing algorithm to get depth image of geometry model on the server. The depth image contained rendering data such as view matrix, color and depth buffers, which were sent back to client mobile devices for rendering with a wrapping equation \cite{McMillan95}. It could run about 6 fps on a 206MHz StrongArm mobile processor. Fritz et al. \cite{Fritz06} conducted object detection and recognition with a modified SIFT algorithm on server to search database for object information descriptor, which was sent back to client for annotation. It run 8 times faster with 98\% accuracy than original implementation. Chen et al. \cite{Chen09} streamed live videos on mobile phone to remote server, on which a SURF-based recognition engine was used to obtain features. It took only 1.0 second recognition latency to detect 150 SURF features from each 320*240 frame. Gu et al. \cite{Gu11} conducted marker detection on server and processed graphics rendering on mobile client. Captured live video was compressed with YUV420 format to reduce data transmission. \par
\emph{Remote-rendering remote-computing (R3C)}: computational and rendering tasks are both outsourced to remote servers and client mobile devices are used as display and user interface. Pasman and Woodward \cite{Pasman03} imported the ARToolkit onto remote server for marker tracking. Virtual overlays were blended with captured images on server side, which were encoded and sent back to client for display. It took about 0.8s to visualize a 3D model with 60,000 polygons on a PDA ten years ago. Lamberti et al. \cite{Lamberti03} visualized 3D scene on remote clusters by using Chromium system to split computational task. Results were reassembled and sent back to client PDA as still images stream. It could display complex models realistically at interactive frame rate. They \cite{Lamberti07} further transmit video stream to server. The server could tailor results according to screen resolution and bandwidth to guarantee realtime performance. With fast development of wireless network and cloud computing technologies, a few works imported computational and rendering tasks into cloud to gain performance improvement. Luo \cite{Luo09} proposed conception of Cloud-Mobile Convergence for Virtual Reality (CMCVR) to improve VR/AR system based on cloud-based infrastructure. He implemented a complicated vision-based gesture interface based on original system. Runtime performance was almost not impacted with help of cloud computing. Lu et al. \cite{Lu11} outsourced whole task on cloud. User input was projected to cloud and rendered screen was compressed and sent back to mobile devices for display.\par
\vspace{2ex} In R3C, as only rendered images are required for display on client mobile devices, transmitted data amount is irrelevant to complexity of virtual scene. In addition, it is free of application migration and compatible problems because all tasks are completed on server and cloud. Both LRRC and R3C suffer from several common problems. The performance is limited by network characteristics including shielding, bandwidth and latency. Data compression alleviates bandwidth traffic to a certain extent at the price of additional computations. In \cite{Wang04, Wolski08}, several principles in terms of bandwidth and computational workload were proposed to guide usage of computation offloading. Comparing to local computation, remote computation also has privacy security problem.
\subsection{Energy Efficiency}
Survey \cite{Paczkowski09} has revealed that battery life is one of the most concerned problems for handheld users. It is especially emphasized for MAR application due to their power-hungry nature. Energy consumption can be addressed at various levels from hardware platform, sensor, network and tracking algorithms to user interaction. Rodriguez and Crowcroft \cite{Rodriguez13} gave a detailed survey on energy management for mobile devices. Most methods mentioned are also suitable for MAR applications. In this section we investigate several power saving technologies not mentioned in the survey, but may be useful for MAR applications.\par
	Semiconductor principle proves that power consumption is exponential to frequency and voltage. Single core CPU improves computational performance with an exponential jump in power consumption, whereas multicore CPU improves performance at the cost of linear increment in power consumption attributed to the fact that each core run at low frequency and voltage when workload is allocated to multiple cores. Nvidia \cite{NVIDIA10} showed a 40\% power improvement by using multicore low-frequency CPU. High parallelism at low clock frequency is much more energy efficient than low parallelism at high clock frequency. MAR applications can also benefits energy saving from leveraging multicore CPU for performance acceleration .\par
	Energy consumption is also proportional to memory access. It is an efficient way to reduce energy consumption by limiting memory access and bandwidth traffic. Arnau et al. \cite{Arnau12} decoupled memory access from fragment calculation to obtain 34\% energy saving. Aforementioned bandwidth saving technologies, such as PCU \cite{Hasselgren07}, TBR \cite{PowerVR11} and data compression \cite{Moller08, Capin08}, also reduce energy consumption as long as energy reduction is greater than energy exhausted by additional operations. Sohn et al. \cite{Sohn05} designed a low-power 3D mobile graphics processor by dynamic configuration of memory according to bandwidth requirements. It reduced power consumption by partly activation of local memory to meet 3D rendering operations.\par
	We can save energy by putting task and hardware components that do not work to sleep to . Clock gating \cite{Moller08} is widely used in the design of mobile devices for energy saving from circuit level. Mochocki et al. \cite{Mochocki06} employed dynamic voltage and frequency scaling (DVFS) technology to optimize mobile 3D rendering pipeline based on previous analytical results. The method was able to reduce power consumption by 40\%. For client-server based applications, the network interfaces are idle in most time to wait for data. It was scheduled to work asynchronously to interleave with other tasks so as to save energy consumption \cite{Wagner09b}.\par
	Computation offloading also conditionally improves energy efficiency. Kumar et al. \cite{Kumar10} proposed a model to evaluate energy consumption by using computation offloading methods. It indicated that energy improvement depends on computations, wireless bandwidth and data amount to transmit through network. System also benefits from energy efficiency if a large amount of computation is offloaded with limited data to transmit. Kosta et al. \cite{Kosta12} proposed a ThinkAir framework to help developers migrate mobile applications to the cloud. The framework provided method-level computation offloading. It used multiple virtual machine (VM) images to parallelize method execution. Since MAR is typical computationally intensive and current networks bandwidth is relative high, most MAR applications obtain energy benefit from computation offloading.
\subsection{Performance vs. Energy Efficiency}
As mobile processors are usually designed with emphasis on lower power consumption rather than performance \cite{Cheng11}, energy efficiency is even more important than performance on mobile devices. In certain conditions, performance promotion also improves energy efficiency, whereas in other cases it increases energy consumption. If we improve performance with operation reduction such as decreasing data amount and memory access \cite{Moller08, Capin08, PowerVR11}, it will also decrease power consumption; if we improve performance by using more powerful hardware \cite{Herling10, Hofmann12}, it will increases energy consumption. In most cases ``increasing'' method obtain higher performance gain than ``decreasing'' way with the cost of higher energy consumption. A more practical solution is to combine them together by fully using available computing resource with the consideration of energy saving.
\section{Systems and applications}
\subsection{Application Fields}
\subsubsection{Tourism and Navigation}
Researchers at Columbia University built a MAR prototype for campus exploration \cite{Hollerer01, Hollerer04}. It overlaid virtual information on items in visitor's vicinity when they walked around campus. Dahne et al. \cite{Dahne02} developed Archeoguide to provide tourists with interactive personalized information about historical sites of Olympia. Tourists could view computer-generated ancient structure at its original site. Fockler et al. \cite{Fockler05} presented an enhanced museum guidance system PhoneGuide to introduce exhibitions. The system displayed additional information on mobile phones when visitors targeted their mobile phones at exhibits. They implemented a perception neuronal network algorithm to recognize exhibits on mobile phone. Elmqvist et al. \cite{Elmqvist06} proposed a three-dimensional virtual navigation application to support both path finding and highlighting of local features. A hand gesture interface was developed to facilitate interaction with virtual world. Schmalstieg et al. \cite{Schmalstieg07b} constructed a MAR-based city navigation system. The system supported path finding and real object selection. When users selected a item in real world, virtual information was overlaid on view of real items. Tokusho and Feiner \cite{Tokusho09} developed a ``AR street view" system which provided an intuitive way to obtain surrounding geographic information for navigation. When users walked on a street, street name, virtual paths and current location were overlaid on real world to give users a quick overview of environment around them. Radha and Diptiman \cite{Radha12} proposed to integrate mobile travel planner with social networks using MAR technologies to see feedbacks and share experiences.
\subsubsection{Entertainment and Advertisement}
Renevier and Nigay \cite{Renevier01} developed a collaborative game based on Studierstube. The game supported two users to play virtual 3D chess on a real table. Piekarski and Thomas \cite{Piekarski02} proposed an outdoor single-player MAR games ARQuake. It enabled players to kill virtual 3D monsters with physical props. Cheok et al. \cite{Cheok03} developed an MAR game for multiplayers in outdoor environment. They embedded Bluetooth into objects to associate them with virtual objects. System could recognize and interact with virtual counterparts when players interacted with real objects. Players killed virtual enemies by physical touch of other players in real world. Henrysson et al. \cite{Henrysson05b} built a face-to-face collaborative 3D tennis game on Symbian mobile phones. Mobile phone was used as tangible racket to hit virtual ball. Hakkarainen and Woodward \cite{Hakkarainen05} created a similar table tennis MAR game. They proposed a color-based feature tracking approach to improve runtime performance. The game was feasible and scalable in various conditions as it required no pre-defined markers. Morrison et al. \cite{Morrison09} developed a MAR map MapLens to superimposed virtual contents on paper map when users targeted mobile phones at paper map. The system was used to help players complete their game. There are also many MAR applications exploited to augment advertisements. The website \cite{MARAd} listed several MAR-based apps for advertisement. Absolut displayed augmented animation to show consumers how vodka was made as users pointed their mobile phones at a bottle of vodka. Starbucks enabled customers to augment their coffee cups with virtual animation scenes. Juniper research [Juniper09] estimated that MAR-based mobile apps and mobile advertising market would reach \$732 million in 2014.
\subsubsection{Training and Education}
The Naval Research Lab (NRL) developed a MAR-based military training system \cite{Julier00} to train soldiers for military operations in different environments. The battlefield was augmented with virtual 3D goals and hazards which could be deployed beforehand or dynastically at run time. Traskback and Haller \cite{Traskback04} used MAR technology to augment oil refinery training. Traditional training was conducted in classrooms or on-site when the plant was shut down for safety consideration. The system enabled on-site device-running training so that trainees could look into run-time workflow. Klopfer et al. \cite{Klopfer05} proposed a collaborative MAR education system for museum. Players used Pocket PCs and walkie-talkies to interview virtual characters and operate virtual instruments. To finish the task, children were encouraged to engage in exhibits more deeply and broadly. Schmalstieg and Wagner \cite{Schmalstieg07a} built a MAR game engine based on Studierstube ES. In order to finish the task, players had to search around with cues displayed on handheld device when players pointed their handheld device at exhibits. Freitas and Campos \cite{Freitas08} developed an education system SMART for low-grade students. Virtual 3D models such as cars and airplanes were overlaid on real time video to demonstrate concepts of transportation and animations.
\subsubsection{Geometry Modeling and Scene Construction}
Baillot et al. \cite{Baillot01} developed a MAR-based authoring tools to create geometry models. Modelers extracted key points from real objects and then constructed geometric primitives from points to create 3D models. Created models were registered and aligned with real objects for checking and verification. Piekarski and Thomas \cite{Piekarski03b} built a similar system for outdoor objects creation. It used pinch gloves and hand tracking technologies to manipulate models. The system was specially suitable for geometrical model creation of giant objects (e.g. building) as users could stand a distance away. Ledermann et al. \cite{Ledermann05} developed a high-level authoring tool APRIL to design MAR presentation. They integrated it into a furniture design application. Users could design and construct virtual model with real furniture as reference in the same view. Henrysson et al. \cite{Henrysson05a} employed MAR to construct 3D scene on mobile phone in a novel way. Motions of mobile phone were tracked and interpreted to translation and rotation manipulation of virtual objects. Bergig et al. \cite{Bergig09} developed a 3D sketching system to create virtual scenes for augmented mechanical experiments. Users used their hands to design experiment scene superimposed on a real drawing. Hagbi et al. \cite{Hagbi10} extended it to support virtual scene construction for augmented games.
\subsubsection{Assembly and Maintenance}
Klinker et al. \cite{Klinker01} developed a MAR application for nuclear plant maintenance. The system created an information model based on legacy paper documents so that users could easily obtain related information that was overlaid on real devices. The system was able to highlight fault devices and supplied instructions to repair them. Billinghurst et al. \cite{Billinghurst08} created a mobile phone system to offer users step-by-step guidance for assembly. A virtual view of next step was overlaid on current view to help users decided which component to add and where to place it in the next step. Henderson and Feiner \cite{Henderson11} developed a MAR-based assembly system. Auxiliary information such as virtual arrows, labels and aligning dash lines were overlaid on current view to facilitate maintenance. A study case showed that users completed task significant faster and more accurate than looking up guidebooks. An empirical study \cite{Tang03} showed that MAR helped to reduced assembly error by 82\%. In addition, it decreased mental effort for users. However, how to balance user attention between real world and virtual contents to avoid distraction due to over-reliance is still an open problem.
\subsubsection{Information Assistant Management}
Goose et al. \cite{Goose04} developed a MAR-based industrial service system to check equipment status. The system used tagged visual markers to obtain identification information, which was sent to management software for equipment state information. Data such as pressure and temperature were sent back and overlaid for display on the PDA. White et al. \cite{White06} developed a head-worn based MAR system to facilitate management of specimens for botanists in the field. The system searched a species database and listed related species samples side-by-side with physical specimens for comparison and identification. Users slid virtual voucher list with head horizontal rotation and zoomed the virtual voucher by head nodding movements. Deffeyes \cite{Deffeyes11} implemented an Asset Assistant Augmented Reality system to help data center administrators find and manage assets. QR code was used to recognize asset. Asset information was retrieved from a MAMEO server and overlaid on current view of asset.\par
\vspace{2ex} MAR also finds its markets in other fields. MAR has been used to enhance visualization and plan operations by placing augmented graphic scans over surgeons' vision field \cite{Wacker06}. Rosenthal et al. \cite{Rosenthal02} used the "x-ray vision" feature to look through the body and made sure the needle was inserted at the right place. MAR was also used to manage personal information \cite{GoogleGlass}. Another large part of applications is AR browsers on mobile phones \cite{Layer, Wikitude}. AR browser is similar to MAR navigation application, but more emphasizes on location-based service (LBS). Grubert et al. \cite{Grubert11} conducted a detailed survey about AR browsers on mobile phones.

\subsection{Representative Systems}
\subsubsection{MARS}
MARS \cite{Hollerer01, Hollerer04} is a both indoor and outdoor MAR system developed by a team at Columbia University. They have built several iterations of the prototype and developed a series of hardware and software infrastructures. The system comprises a backpack laptop, a see-through head-worn display and several input devices including stylus and trackpad. An orientation tracker and RTK GPS are used to obtain pose information. A campus guide system has been developed based on MARS. Visitors are able to obtain detailed information overlaid on items in their current view field. They can also watch demolished virtual buildings on their original sites. It supports virtual menus overlaid on users view field to conduct different tasks. The system can also be used for other applications such as tourism, journalism, military training and wayfinding.
\subsubsection{ARQuake}
ARQuake \cite{Piekarski02} is a single-player outdoor MAR games based on popular desktop game Quake. Virtual monsters are overlaid on current view of real world. Player moves around real world and uses real props and metaphors to kill virtual monsters. Real buildings are modeled but not rendered for view occlusion only. The game adopts GPS and digital compass to track player's position and orientation. A vision-based tracking method is used for indoor environments. As virtual objects may be difficult to recognize from natural environments at outdoor environments, system have to run several times to set a distinguishable color configuration for later use.
\subsubsection{BARS}
BARS \cite{Julier00} is a battlefield augmented reality system for soldiers training in urban environments. Soldiers' perceptions of battlefield environment are augmented by overlaying building, enemies and companies locations on current field of view. Wireframe plan is superimposed on real building to show its interior structures. An icon is used to report location of sniper for threat warning or collaborative attacking. A connection and database manager is employed for data distribution in an efficient way. Each object is created at remote servers but only simplified ghost copy is used on clients to reduce bandwidth traffic. The system requires a two-step calibration. The first is to calculate mapping of result from a sensing device to real position and orientation of sensors; the second is to map sensing unit referential to viewpoint referential of user's observation.
\subsubsection{Medien.welten}
Medien.welten \cite{Schmalstieg07a} is a MAR system that has been deployed at Technisches Museum Wien in Vienna. The system is developed based on Studierstube ES. A scene graph is designed to facilitate construction and organization of 3D scenes. Total memory footprint is limited to 500k to meet severe hardware limitation on handheld devices. Game logic and state are stored in a XML database in case of client failure due to wireless single shielding. Medien.welten enables players to use augmented virtual interface on handheld devices to manipulate and communicate with real exhibits. With interactive manipulation, players gain both fun experience and knowledge of exhibits.
\subsubsection{MapLens}
MapLens \cite{Morrison09, Morrison11} is an augmented paper map. The system employed mobile phones' viewfinder, or ``magic lens", to augment paper map with geographical information. When users view a paper map through embedded camera, feature points on paper map are tracked and matched against feature points tagged with geographical information to obtain GPS coordinates and pose information. GPS coordinates are used to search an online HyperMedia database (HMDB) to retrieve location-based media such as photos and other metadata, which are overlaid on paper map from current pose. Augmented maps can be used in collaborative systems. Users can share GPS-tagged photos with others by uploading images to HMDB so that others can view new information. It establishes a common ground for multiple users to negotiate and discuss to solve the task in a collaborative way. Results show that it is more efficient than digital maps.
\subsubsection{Virtual LEGO}
Virtual LEGO \cite{Henrysson05a} uses mobile phone to manipulate virtual graphics objects for 3D scene creation. The motion of mobile phone is employed to interact with virtual objects. Virtual objects are fixed relative to mobile phone. When users move their mobile phones, objects are also moved according to relative movement of mobile phones to the real world. In translation mode, the selected object is translated by the same distance as mobile phone. Translation of mobile phone is projected onto a virtual Arcball and converted as rotation direction and angle to rotate virtual object. The objects are organized in a hierarchical structure so that transformation of a parent object can be propagated to its sub-objects. A multiple visual markers tracking method is employed to guarantee accuracy and robustness of mobile phone tracking. Result shows that the manipulation is more efficient than button interface such as keypad and joypad, albeit with relative low accuracy.
\subsubsection{InfoSPOT}
InfoSPOT \cite{Irizarry12} is a MAR system to help facility managers (FMs) access building information. It augments FMs' situation awareness by overlaying device information on view of real environment. It enables FMs to fast solve problems and make critical decisions in their inspection activities. The Building Information Modeling (BIM) model is parsed into geometry and data parts, which are linked with unique identifiers. Geo-reference points are surveyed beforehand to obtain accurate initial registration and indoor localization. The geometry part is used to render panoramas of specific locales to reduce sensor drift and latency. When FMs click virtual icon of physical object on IPad screen, its identifier is extracted to search data model to fetch information such as product manufacture, installation date and life of product.
\subsubsection{Google Glass}
Google Glass \cite{GoogleGlass} is a wearable AR device developed by Google. It displays information on glass surface in front of users' eyes and enables users to control interface with natural language voice commands. Google Glass supports several native functions of mobile phones such as sending messages, taking pictures, recording video, information searching and navigation. Videos and images can be shared with others through Google+. Current product uses smartphones as network transition for Internet access. As it only focuses on text and image based augmentation on a tangible interface, it does not require tracking and alignments of virtual and real objects. Presently it is only available for developers and reported to be widely delivered later this year.
\subsubsection{Wikitude}
Wikitude \cite{Wikitude} is a LBS-based AR browser to augment information on mobile phones. It is referred as ``AR browser'' due to its characteristic of augmentation with web-based information. Wikitude overlays text and image information on current view when users point their mobile phones to geo-located sites. Wikitude combines GPS and digital compass sensors to track pose tracking. Contents are organized in KML and ARML formats to support geographic annotation and visualization. Users can also register custom web services to get specific information. \par
\vspace{2ex} Table~\ref{tab:five} lists system components and enabling technologies of aforementioned MAR applications. Early applications employed backpack notebook computer for computing tasks. External HMDs were required to provide optical see-through display. As mobile devices become powerful and popular, numerous applications use mobile devices as computing platforms. Embedded camera and self-contained screen are used for video see-through display. Single tracking methods have also been replaced with hybrid methods to obtain high accurate results in both indoor and outdoor environments. Recently applications outsource computations on server and cloud to gain acceleration and reduce client mobile device requirements. With rapid advances from all aspects, MAR will be widely used in more application fields.
 \begin{table}%
\tbl{Several representative MAR applications\label{tab:five}}{%
\begin{tabular*}{\textwidth}{|p{1.0cm} |p{1.2cm}| p{1.2cm}| p{1.0cm}| p{1.2cm}| p{1.0cm}| p{1.3cm}| p{1.0cm}| p{1.25cm}|}
\hline
System  & Hardware platform   & Software platform   & Display    & Tracking   & Network & Collaborative   & Indoor/ outdoor & Application field\\\hline
MARS	& notebook computer	& JABAR/ Ruby	&optical see-through HMD	& RTK GPS, orientation tracker	& campus WLAN	& multiusers 	 &indoor, outdoor	& tourism, navigation \\\hline
ARQuake	& notebook computer	& Tinmith-evo5	&optical see-through HMD	& GPS, digital compass	& cable network	& single 	 &indoor, outdoor	& games \\\hline
BARS	& notebook computer	& unknown	&optical see-through HMD	& inertial sensors, GPS	& WWAN	& multiusers 	&outdoor	& training \\\hline
Medien. welten & PDA	& Studiers- tube ES	&video see-through screen	& visual marker tracking & WLAN	& multiusers 	&indoor	& education, games \\\hline
MapLens	& mobile phone	& unknown	&video see-through phone screen	& nature feature tracking	& WWAN	& multiusers 	&indoor, outdoor	& tourism, game \\\hline
Virtual LEGO	& mobile phone	& UMAR	&video see-through phone screen	& visual marker tracking	& unknown	& single 	 &indoor	& authoring, assembly \\\hline
InfoSPOT	& Apple iPad	& KHARMA	&video see-through pad screen	& sensors, geo-reference visual markers	& WWAN	& single 	 &indoor	& information management  \\\hline
Google Glass	& AR glass	& unknown	&optical see-through glass	& unknown 	& WPAN + WWAN	& multiusers 	&indoor	& information management, navigation  \\\hline
Wikitude	& mobile phone	& Wikitude SDK	&video see-through phone screen	& GPS, sensors, visual markers & WWAN, Wi-Fi	& multiusers 	&indoor, outdoor	& information management, navigation  \\\hline
\end{tabular*}}
\begin{tabnote}%
\vskip2pt
\end{tabnote}%
\end{table}%
\section{Challenging problems}
\subsection{Technology limitations}
MAR develops based on various technologies as mentioned above. Many problems such as network QoS deficiency and display limitations remain unsolved in their own fields. Some problems are induced by the combination of multiple technologies. For instance, battery capacity is designed to be sustainable for common functions such as picture capturing and Internet access, which are supposed to be used at intervals. MAR applications require long-time cooperation of cameras capturing, GPS receiving and Internet connection. These tasks working together can drain battery quickly. Many high accurate tracking approaches are available in computer vision fields but they can not be directly used on mobile devices due to limited computing capability. We have discussed technology related challenges in previous sections. Several papers \cite{Azuma01, Hollerer04, Papagiannakis08} also made detailed investigation of it. MAR also has several intrinsic problems from technology aspect, which are very much underexplored but worth great consideration. We will address these challenges in the following sections. \par
\subsection{Privacy and security}
Privacy and security are especially serious for MAR due to various potential invasion sources including personal identification, location tracking and private data storage. Many MAR applications depend on personal location information to provide services. For instance, in client-server applications, user's position is transmitted to third-party server for tracking and analysis, which may be collected over time to trace user activity. It is more serious for collaborative MAR applications as users have to share information with others, which not only provides opportunity for others to snoop around private information but also raise concern of how to trust quality and authenticity of user-generated information supplied by others. To guarantee privacy safety, we require both trust models for data generation and certification mechanisms for data access. Google Glass is a typical example to show users' concern about privacy. Although it is not widely delivered yet, it has already raised privacy concern that users can identify strangers by using facial recognition or surreptitiously record and broadcast private conversations. Acquisti et al. \cite{Acquisti12} discussed privacy problem of facial recognition for AR applications. Their experiment implied that users could obtain inferable sensitive information by face matching against facial images from online sources such as social network sites (SNS). They proposed several privacy guidelines including openness, use limitation, purpose specification and collection limitation to protect privacy use.\par
\subsection{Application breakthrough}
Most existing MAR applications are only prototypes for experiment and demonstration purposes. MAR has great potential to change our ways to interact with real world, but it still lacks killer applications to show its capability, which may make it less attractive for most users. Breakthrough applications are more likely to provide a way for drivers to see through buildings to avoid cars coming from cross streets or help backhoe operator to watch out fiber-optic cables buried underground in the field. We have witness similar experience for Virtual Reality (VR) development during past decades, so we should create feasible applications to avoid risk of the same damage to MAR as seen when VR was turned from hype to oblivion. Google Glass is a milestone product to raise public interest but it is still stuck with absence of killer applications. Google has delivered its explorer edition for developers so as to create several significant applications before it is widely available for public users.\par
\subsection{Over-emphasized self-support}
Many MAR systems are designed to be self-contained to make it free from environmental support. Self-support is emphasized to map completely unknown surroundings and improve user experience. However, it introduces complexity and limitations. For instance, many systems employ visual feature method to get rid of beforehand deployed visual markers, but deficiencies of heavy computational overhead and poor robustness make the system even less applicable for most applications. Besides, what useful annotations can be expected if we know nothing about the environment? It is still unclear about the necessity to make it universal for completely unprepared surroundings. With development of pervasive computing and Internet of Things (IOTs), computing and identification are woven into the fabric of daily life and indistinguishable from environments. The system may be deeply integrated with environment other than isolated from it. Another reason to emphasize self-support necessity is for outdoor usage. A study case \cite{LaMarca05} showed that GPS usage coverage was very lower than expected. As GPS was shielded in indoor environments,  it indicated that users may spent most of their time indoors, so there may be not so great urgency to make system completely self-contained.\par
\subsection{Social acceptance}
Many factors such as device intrusion, privacy and safety considerations may affect social acceptance of MAR. To reduce system intrusion, we should both miniaturize computing and display devices and supply a nature interactive interface. Early applications equipped with backpack laptop computer and HMDs introduce serious device intrusion. Progresses in miniaturization and performance of mobile devices alleviate the problem to certain extent, but they do not work in a nature way. Users have to raise their hands to point cameras at real objects during system operation, which may cause physical fatigue for users. The privacy problem is also seriously concerned. A ``Stop the Cyborgs'' movement has attempted to convince people to ban Google Glass in their premises. Many companies also post anti-Google Glass signs in their offices. As MAR distracts users' attention from real world occasionally, it also induce safety problems when users are operating motor vehicles or walking in the streets. All these issues work together to hurdle the social acceptance of MAR technology.\par
\section{Conclusions}
In this paper, we give a complete and detailed survey of MAR technology in terms of system components, enabling technologies and application fields. Although there are still several problems from technical and application aspects, it is estimated as one of the most promising mobile applications. MAR has become an important manner to interact with real world and will change our daily life.\par
As fast development of cloud computing and wireless networks, mobile cloud computing becomes a new trend to combine the high flexibility of mobile devices and the high-performance capabilities of cloud computing. It will play a key role in MAR applications since it can undertake the heavy computational tasks to save energy and extend battery lifetime. Furthermore, cloud services for MAR applications can operate as caches and decrease the computational cost for both MAR devices and cloud providers. As MAR applications run on a remote server, we can also overcome limitations of mobile operating systems with help of mobile browsers. It is possible to combines multiple mobile devices for cooperative mobile computing which will be suitable for collaborative MAR applications such as multi-player games, collaborative design and virtual meeting. Although there are still several problems such as bandwidth limitation, service availability, heterogeneity and security, mobile cloud computing and cooperative mobile computing seem promising new technologies to promote MAR development to a higher level.\par
Present MAR applications are limited to mobile devices like PADs and mobile phones. We believe that these mobile devices are transient choices for MAR as they are not originally designed for MAR purpose. They happen to be suitable but not perfect for it. Only dedicated devices such as Google Glass can fully explore potential capability of MAR. As development of mobile computing and wearable computers such as AR glass and wristwatch devices, we are looking forward to its renaissance and prosperity around the corner.

% Appendix
\appendix
% Bibliography
\small
\bibliographystyle{ACM-Reference-Format-Journals}
\bibliography{acmsmall-sample-bibfile}
                             % Sample .bib file with references that match those in
                             % the 'Specifications Document (V1.5)' as well containing
                             % 'legacy' bibs and bibs with 'alternate codings'.
                             % Gerry Murray - March 2012

% History dates
%\received{February 2007}{March 2009}{June 2009}

% Electronic Appendix
%\elecappendix

%\medskip

%\section{Appendix}

%The primary consumer of energy in WSNs is idle listening. The key to
%reduce idle listening is executing low duty-cycle on nodes. Two
%primary approaches are considered in controlling duty-cycles in the
%MAC layer.

\end{document}